\documentclass[prb,twocolumn,showpacs,preprintnumbers,amsmath,amssymb,superscriptaddress]{revtex4}
\usepackage{amsmath,amssymb,amsfonts} 
\usepackage{graphics}                 

\usepackage[dvips]{graphicx}
\usepackage[latin1]{inputenc}
\usepackage{multirow}

\newcommand{\rrr}{\mathbf{r}}

\begin{document}

\title{Delta Self-Consistent Field as a method to obtain potential energy surfaces of excited molecules on surfaces}

\author{Jeppe Gavnholt}
\affiliation{Danish National Research Foundation's Center for Individual Nanoparticle Functionality (CINF),
             Department of Physics,
             Technical University of Denmark,
             DK--2800 Kongens Lyngby,
             Denmark}%

\author{Thomas Olsen}
\affiliation{Danish National Research Foundation's Center for Individual Nanoparticle Functionality (CINF),
             Department of Physics,
             Technical University of Denmark,
             DK--2800 Kongens Lyngby,
             Denmark}%

\author{Mads Engelund}
\affiliation{Danish National Research Foundation's Center for Individual Nanoparticle Functionality (CINF),
             Department of Physics,
             Technical University of Denmark,
             DK--2800 Kongens Lyngby,
             Denmark}%
\affiliation{\emph{Present address:} Department of Micro- and Nanotechnology,
             Technical University of Denmark,
             DK--2800 Kongens Lyngby,
             Denmark}%

\author{Jakob Schi{\o}tz}
\email{schiotz@fysik.dtu.dk}
\affiliation{Danish National Research Foundation's Center for Individual Nanoparticle Functionality (CINF),
             Department of Physics,
             Technical University of Denmark,
             DK--2800 Kongens Lyngby,
             Denmark}%

\date{\today}

\begin{abstract}
We present a modification of the $\Delta$SCF method of calculating energies of excited states, in order to make it applicable to resonance calculations of molecules adsorbed on metal surfaces, where the molecular orbitals are highly hybridized. The $\Delta$SCF approximation is a density functional method closely resembling standard density functional theory (DFT), the only difference being that in $\Delta$SCF one or more electrons are placed in higher lying Kohn-Sham orbitals, instead of placing all electrons in the lowest possible orbitals as one does when calculating the ground state energy within standard DFT. We extend the $\Delta$SCF method by allowing excited electrons to occupy orbitals which are linear combinations of Kohn-Sham orbitals. With this extra freedom it is possible to place charge locally on adsorbed molecules in the calculations, such that resonance energies can be estimated, which is not possible in traditional $\Delta$SCF because of very delocalized Kohn-Sham orbitals. The method is applied to N$_2$, CO and NO adsorbed on different metallic surfaces and compared to ordinary $\Delta$SCF without our modification, spatially constrained DFT and inverse-photoemission spectroscopy (IPES) measurements. This comparison shows that the modified $\Delta$SCF method gives results in close agreement with experiment, significantly closer than the comparable methods. For N$_2$ adsorbed on ruthenium (0001) we map out a 2-dimensional part of the potential energy surfaces in the ground state and the 2$\pi$-resonance. From this we conclude that an electron hitting the resonance can induce molecular motion, optimally with 1.5 eV transferred to atomic movement. Finally we present some performance test of the $\Delta$SCF approach on gas-phase N$_2$ and CO, in order to compare the results to higher accuracy methods. Here we find that excitation energies are approximated with accuracy close to that of time-dependent density functional theory. Especially we see very good agreement in the minimum shift of the potential energy surfaces in the excited state compared to the ground state.
\end{abstract}
\maketitle

\section{Introduction}

Density functional theory\cite{hohenbergkohn,kohnsham} (DFT) has proved to be a vital tool in gaining information on many gas-surface processes. This may be surprising, since DFT is only valid for relaxed systems in their ground state and therefore not directly applicable to dynamical situations. However, often the electrons relax much faster than the time scale of the atomic movement, such that the electron gas can be considered relaxed in its ground state at all times. Then potential energy surfaces (PES) of the ground state obtained by DFT, or any other method, can be used to describe the motion of atomic cores. This is the Born-Oppenheimer approximation.

In some situations, however, the Born-Oppenheimer approximation is not valid. This is for example the case when the electronic system is excited by a femtosecond laser\cite{femtosecond_Luntz,science_RuCO} or hot electrons are produced with a metal-insulator-metal junction.\cite{gadzuk_1996} The Born-Oppenheimer approximation also breaks down if the time scales for the electronic and nuclear motions are comparable or if the separations between the electronic states are very small, such that transitions between the electronic states will occur. In these situations it is necessary to go beyond the Born-Oppenheimer approximation either by considering the coupling between electronic states\cite{electronic_coupling_Lichten,behler_2007} where it becomes necessary to obtain PESs of excited states, or by an electronic friction model.\cite{Bohnen_friction_1975,Brandbyge_friction_1995}

The problem of calculating excitation energies are being approached in many different ways, even within DFT. Time dependent density functional theory (TDDFT)\cite{runge_gross} gives, compared to the computational cost, good agreement with experiments for excitations in atoms and molecules.\cite{tddft_comp1} However, TDDFT suffers some problems in excitations involving charge transfer.\cite{tddft_charge} The GW approximation\cite{GW_1,gw_rubio} can be used to gain accurate excitation energies for molecules and clusters. The embedding method,\cite{kluner_2002,kluner_2007} which combines high-accuracy quantum chemistry methods with DFT, makes it possible to handle larger periodic systems with great accuracy. The embedding theory has been applied to estimate PESs of excited molecules on surfaces.\cite{embedding_PES} However, the computational cost and involved complexity is still very high. Our aim has been to find a method, which at a computational cost close the level of ground state DFT can estimate excited state energies of molecules on surfaces with reasonable accuracy. Such a method would make it more feasible to consider a large range of systems in search of systems with interesting or desired properties.

Constrained DFT\cite{behler_2007,wu_voorhis_2005,wu_voorhis_2006} and $\Delta$SCF\cite{dscf_1989,hellman} are two different approaches, which both can be considered as small extensions of ground state DFT, such that the computational cost lies close to that of ground state DFT. In constrained DFT an additional potential is introduced and varied until a certain constraint on the electrons is fulfilled. The simplest approach is to lower (or increase) the potential in a certain part of space until you have the desired number electrons in this area.\cite{wu_voorhis_2005} A different approach is to introduce potentials on the orbitals in a localized basis set, which depends on the orbitals' positions in space.\cite{behler_2007} In section \ref{sec:mol_on_surf} we will argue that when considering molecular resonance states on surfaces it may be problematic with such a strict constraint on the electrons, since a part of the charge may return to the surface on a much shorter time-scale than the lifetime of the resonance.

In the $\Delta$SCF scheme the positions of the electrons are controlled by controlling the occupation of the Kohn-Sham (KS) states as the system reaches self-consistency. The $\Delta$SCF scheme has for a long time been justified in cases, where the excited state corresponds to the lowest state of a given symmetry.\cite{dscf_just_1976} The scheme has, however, often been applied to more general cases. More recently, G\"orling\cite{gorling_1999} extended the KS formalism to include excited states, such that $\Delta$SCF gets a formal justification in the general case, although a special unknown orbital-dependent exchange-correlation potential should be used for the excited states. In practical implementations standard exchange-correlation potentials from ground state DFT are typically used.

This traditional way of just controlling the occupation of the KS orbitals has some limitations. For example when a molecule is placed on a metallic surface the molecular orbitals will hybridize with the orbitals in the surface, such that the molecular orbitals will be spread over several KS states. For such systems there is no good way of representing a resonance on the molecule as a change in the occupations of the KS orbitals. The optimal thing one can do within this scheme is to occupy the KS orbital with the largest overlap with the molecular orbital in question, but this overlap can be quite small and highly system size dependent. This problem was also pointed out by Hellman et al.\cite{hellman}\ and Behler et al.\cite{behler_2007}

In this paper we modify the $\Delta$SCF approach, such that electrons are allowed to occupy arbitrary linear combinations of KS orbitals. In this way one achieves much better control on the position of the excited electron. As is the case for traditional $\Delta$SCF some knowledge of the resonance one wants to consider is needed in order to apply the method. The method is especially relevant in Newns-Andersson\cite{newns,andersson} type systems, where a resonance can be attributed to a known single level, which has been hybridized through interactions with other states. This includes systems with molecules adsorbed on metal surfaces and molecules trapped between to metal contacts.

The modification we propose only has minor implications on the way practical calculations are performed, which is very similar to performing an ordinary ground state DFT calculation. In the following we will go through the details of the method and apply it to a few diatomic molecules on metallic surfaces. The obtained results will be compared to the ordinary $\Delta$SCF method, spatially constrained DFT and IPES measurements. Finally we present some tests on the performance of the $\Delta$SCF approach on N$_2$ and CO in the gas phase.

\section{Method}

In the following we go through the differences between the linear expansion $\Delta$SCF method we propose, ordinary $\Delta$SCF and standard DFT. We start by stating the modification of the KS equations when considering an electron excited from the Fermi level to a higher lying state. Then we show how this affects the energy calculation. Finally we generalize the approach to other types of excitations.

\subsection{Kohn-Sham equations}\label{sec:method_KS}

The fundamental KS equations\cite{kohnsham} represent a practical way of finding the ground state electron density for a given external potential and a given number of electrons through an iterative process.
\begin{eqnarray}
\left[-\frac{\nabla^2}{2}+v_{KS}[n](\rrr)\right]\psi_i(\rrr)=\epsilon_i\psi_i(\rrr) \label{eq:KS_s}\\
n(\rrr)=\sum_{i=1}^N\psi^*_i(\rrr)\psi_i(\rrr) \label{eq:KS_n} \\
v_{KS}[n](\rrr)=v_{ext}(\rrr)+\int d\rrr'\frac{n(\rrr')}{\vert \rrr-\rrr' \vert} + \frac{\delta E_{xc}}{\delta n(\rrr)},
\end{eqnarray}
where $v_{KS}$ is the KS potential, $E_{xc}$ is the exchange-correlation energy and $N$ is the number of electrons. As seen from Eq.\ (\ref{eq:KS_n}) only the $N$ orbitals with lowest energy contribute to the density, ie.\ the electrons are placed in these orbitals.\cite{e_temp_footnote} In ordinary $\Delta$SCF one estimates properties of excited states by placing the electrons differently. For example the HOMO-LUMO gap in a molecule could be estimated by replacing Eq.\ (\ref{eq:KS_n}) with
\begin{eqnarray}
n(\rrr)=\sum_{i=1}^{N-1}\psi^*_i(\rrr)\psi_i(\rrr) + \psi^*_a(\rrr)\psi_a(\rrr),
\end{eqnarray}
where $\psi_a(\rrr)$ is the KS orbital resembling the LUMO from the ground-state calculation. Naturally, the KS orbitals found when solving these modified KS equations will differ from the ones found in an ordinary DFT calculation, due to the change in the Hamilton through the change in the density when different orbitals are occupied.

In the linear expansion $\Delta$SCF method we propose, the excited electron is not forced to occupy a KS orbital, but can occupy any orbital that is a linear combination of empty KS orbitals:
\begin{eqnarray}
\psi^{res}(\rrr)=\sum_{i=N}^Ma_i\psi_i(\rrr),\label{eq:psi_res}
\end{eqnarray}
where $M$ is the number of KS orbitals in the calculation. In practice this means that the KS many-particle wavefunction is no longer just a Slater determinant of $N$ KS orbitals, but a Slater determinant of $N-1$ KS orbitals and $\psi^{res}(\rrr)$. Only empty KS orbitals are included in the linear expansion, since otherwise $\psi^{res}(\rrr)$ will not be orthogonal to the filled KS orbitals. Eq.\ (\ref{eq:KS_n}) is then replaced with
\begin{eqnarray}
n(\rrr)&=&\sum_{i=1}^{N-1}\psi^*_i(\rrr)\psi_i(\rrr) + \sum_{i,j=N}^Ma_i^*a_j\psi^*_i(\rrr)\psi_j(\rrr).\label{eq:dscf_n}
\end{eqnarray}
Since the expansion coefficients, $a_i$, in principle could have any value some a priori knowledge is needed in order to choose good values. In the case of molecular resonances on surfaces the expansion coefficients are chosen such that $\psi^{res}(\rrr)$ resembles the relevant molecular orbital as much as possible, ie. 
\begin{equation}
a_i=\frac{\langle\psi_i\vert\phi\rangle}{\left(\sum_i \vert\langle\psi_i\vert\phi\rangle\vert^2\right)^\frac{1}{2}},
\end{equation}
 where $\phi$ is the molecular orbital. This is consistent with a Newns-Andersson\cite{newns,andersson} picture, where the resonance corresponds to an electron getting in the molecular orbital, but the resonance broadening and energy shift is due to hybridization with the metallic bands and an image charge effect.

In calculations with k-point sampling the linear expansion is performed independently in all k-points. In the linear expansion $\Delta$SCF one then avoids the difficulties one can encounter in choosing which KS state to occupy in each k-point in the traditional way of performing $\Delta$SCF calculations: For example one may risk occupying different bands in each k-point, when just choosing the KS orbital with the largest overlap with the molecular orbital in each k-point.

\subsection{The energy}\label{sec:method_energy}

The energy calculation, which is performed after the KS equations have reached self consistency, is not significantly different in the linear expansion $\Delta$SCF scheme compared to ordinary DFT. The Hartree energy is evaluated directly from the density, which is also the case for the exchange-correlation energy if an orbital independent functional is used. So in linear expansion $\Delta$SCF these terms are evaluated exactly as in ordinary DFT. In ordinary DFT the kinetic energy is evaluated as
\begin{eqnarray}
T[n(\rrr)]&=& \sum_{i=1}^N\langle\psi_i\vert-\frac{\nabla^2}{2}\vert\psi_i\rangle \nonumber\\
&=&\sum_{i=1}^N\epsilon_i - \int v_{KS}[n](\rrr)n(\rrr)d\rrr
\end{eqnarray}
where the last equality is seen directly from Eq.\ (\ref{eq:KS_s}). Similarly the expression for the kinetic energy in the linear expansion $\Delta$SCF is found to be
\begin{eqnarray}
T[n(\rrr)]&=&\sum_{i=1}^{N-1}\epsilon_i + \sum_{i=N}^{M}\vert a_i\vert^2\epsilon_i -\nonumber\\
&& \int v_{KS}[n](\rrr)n(\rrr)d\rrr \label{eq:dscf_T}
\end{eqnarray}
For orbital dependent exchange-correlation functionals some effort must be put into ensuring that the exchange-correlation energy is evaluated correctly. This should however be quite straightforward, since all the occupied orbitals are known.

\subsection{Gradients}\label{sec:method_forces}

Gradients of PESs are easily evaluated in ordinary DFT due to the Hellman-Feynman theorem. The Hellman-Feynman theorem, however, only applies to eigenstates and not linear expansions of eigenstates. Due to this there is no easy way of gaining the gradients in a linear expansion $\Delta$SCF calculation. In section \ref{sec:gradients_calc} we will show that the Hellman-Feynman gradients do in fact not match the true gradients.

\subsection{Other excitations}

Above we only considered excitations where an electron is removed from the Fermi energy and placed in some specified orbital. The method is, however, easily extended to other types of excitations by representing each removed and each added electron as a linear expansion of KS orbitals. Eq.\ (\ref{eq:dscf_n}) then gains an extra sum for each extra linear expansion. In cases of removed electrons the sign should of course be negative and the sum be over KS states below the Fermi energy. Similarly Eq.\ (\ref{eq:dscf_T}) gains extra sums.

\subsection{Implementation}

We have implemented the method in \texttt{gpaw},\cite{gpaw_article,gpaw} which is a real-space DFT code that uses the projector-augmented waves\cite{paw1,paw2} (PAW) formalism to represent the core electrons. The self-consistent electron density is determined by an iterative diagonalization of the KS Hamiltonian and Pulay mixing of the resulting density.\cite{kresse96} For calculations on single molecules we use the local density approximation (LDA)\cite{Ceperley_LDA} as well as RPBE\cite{rpbe} to describe exchange and correlation effects. The LDA is used because we compare to TDDFT results obtained using the ALDA approximation.\cite{Zangwill_ALDA} and RPBE is used to see whether or not the generalized gradient description improves results. For calculations on molecules at surfaces we only use RPBE, because this is designed to perform well for molecules adsorbed on transition metal surfaces.

The projection step described in section \ref{sec:method_KS} can easily be approximated within the PAW formalism if the atomic orbitals are chosen as partial waves; see appendix \ref{app:paw} for details.

For reasons of comparison we have also made a few linear response TDDFT (lrTDDFT) calculations. These have been made using the \texttt{Octopus} code,\cite{octopus,octopus_paper} which is a real-space TDDFT code using norm-conserving pseudopotentials to represent core electrons.

\section{Molecules on surfaces}\label{sec:mol_on_surf}

The linear expansion $\Delta$SCF method is especially relevant for molecules on metallic surfaces, because the molecular state, due to hybridization, is spread over many KS states, ie.\ it is necessary to write the resonant state as a linear combination of KS states. In this section we will make a detailed investigation of the 2$\pi$ resonance of N$_2$ on a ruthenium (0001) surface. Furthermore we apply the proposed method to several diatomic molecules on different metallic surfaces and compare the results to other methods and experiments. Finally we map out a part of the PESs for N$_2$ on ruthenium (0001) and use it to estimate how much energy could possible be put into molecular motion from an electron hitting the resonance.

\subsection{$2\pi$ resonance energy for N$_2$ on ruthenium}

\begin{figure}[t]
\begin{center}
\includegraphics[width=0.45\textwidth]{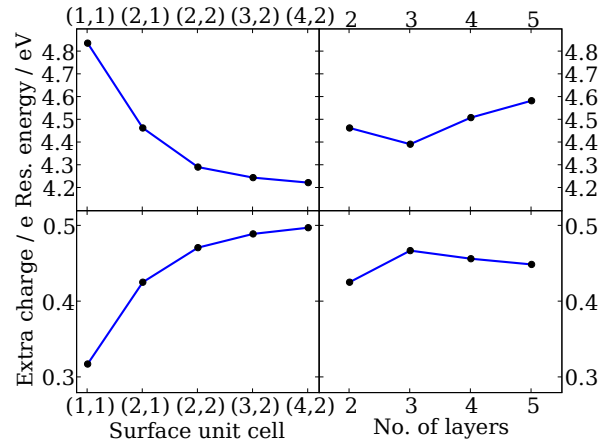}
\caption{(color online) Upper row: The $2\pi$ resonance energy of N$_2$ molecule on a ruthenium surface. Lower row: The extra charge on the N$_2$ molecule in the resonance compared to a ground state calculation. Left panels are for two layers and different surface cells, ie.\ different N$_2$ coverages. Right panels are for a (2,1) surface cell and different number of layers. The extra amount of charge is estimated using Bader decomposition.\cite{bader,bader_book}}\label{fig:bader_convergence}
\end{center}
\end{figure}

The two top panels on Fig.\ \ref{fig:bader_convergence} shows the $2\pi$ resonance energy for N$_2$ on a ruthenium (0001) surface as a function of the system size, ie.\ the surface unit cell and the number of ruthenium layers.
The resonance energy is the total energy difference between a resonant calculation and a ground state calculation, both performed with atomic positions corresponding to the minimum of the ground state PES, ie. it is vertical resonance energies. We minimize the energy in the ground state calculations by keeping all surface atoms frozen and found that the nitrogen molecule is placed on-top with the two nitrogen atoms placed 2.084 {\AA} and 3.201 {\AA} above the surface. In the resonance calculation the $2\pi_y$ orbital of the N$_2$ molecule has been expanded on all KS states above the Fermi energy. This expansion has been used as $\psi^{res}$ in Eq.\ (\ref{eq:psi_res}). Although an extra electron is placed on the molecule we keep the total number of electrons unchanged, such that the unit cell is neutral. This is reasonable because a charged molecule will form an image charge in the surface, keeping the entire system neutral.

\begin{figure}[t]
\begin{center}
\includegraphics[width=0.2\textwidth]{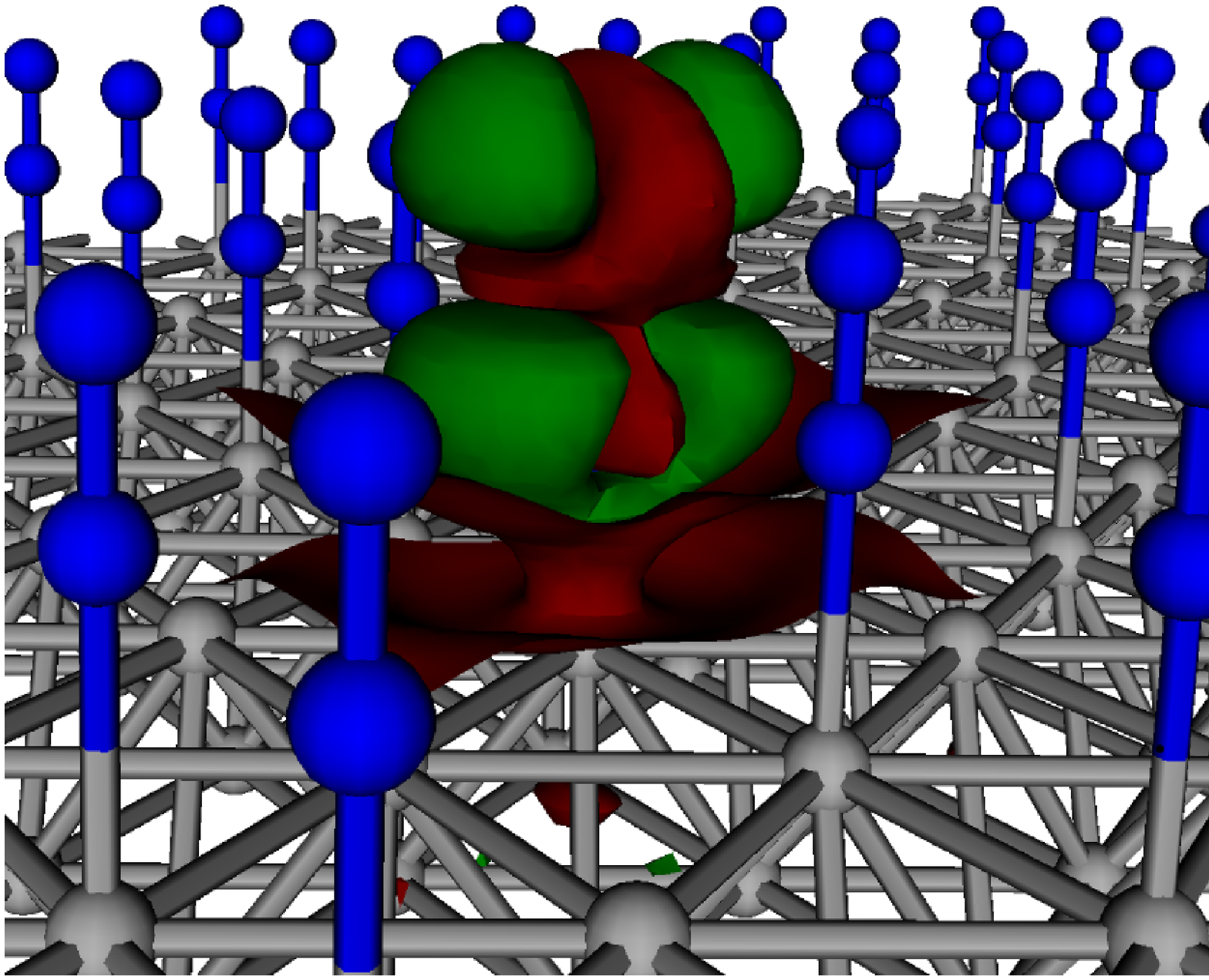}
\includegraphics[width=0.2\textwidth]{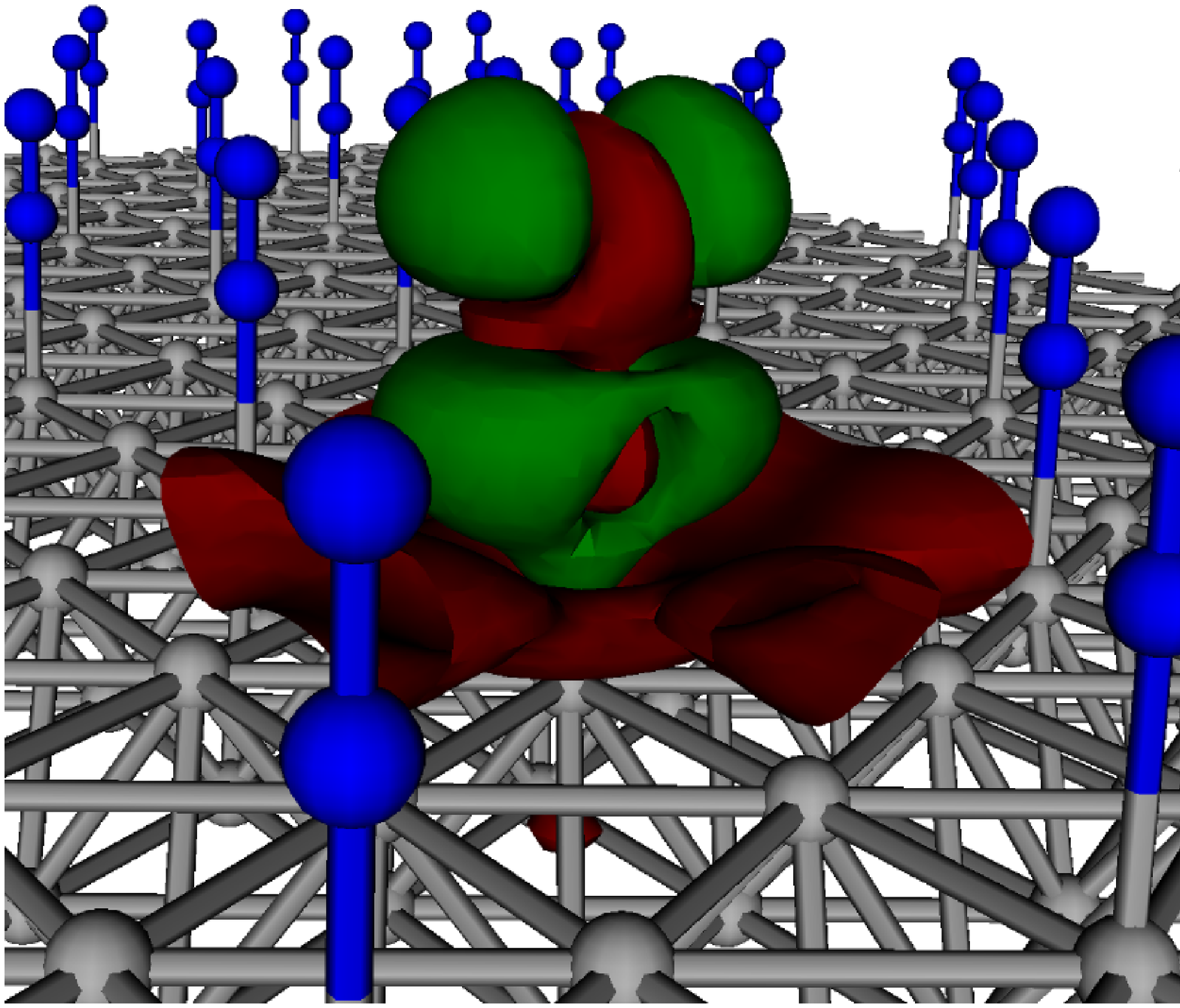}\\
\includegraphics[width=0.2\textwidth]{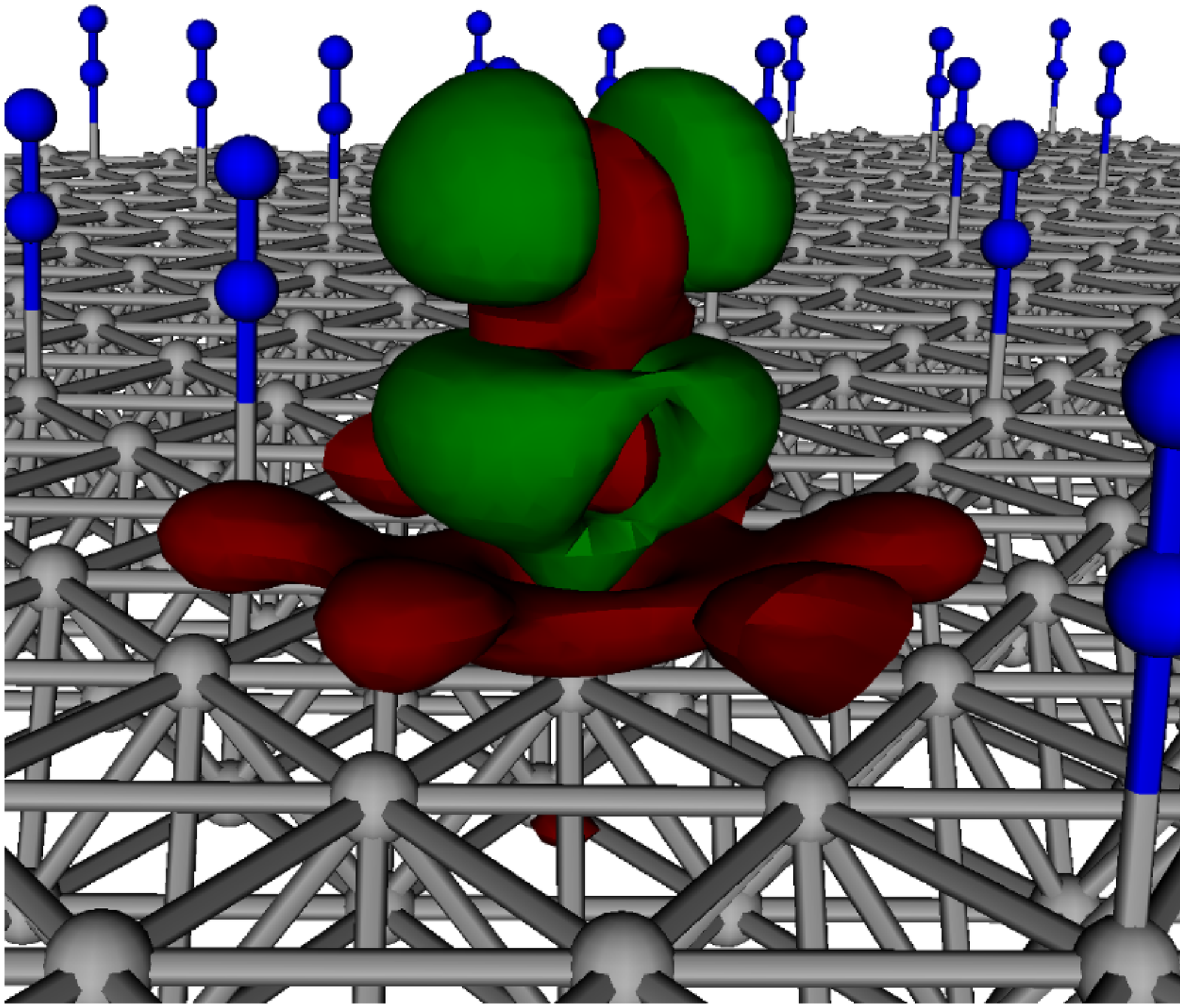}
\includegraphics[width=0.2\textwidth]{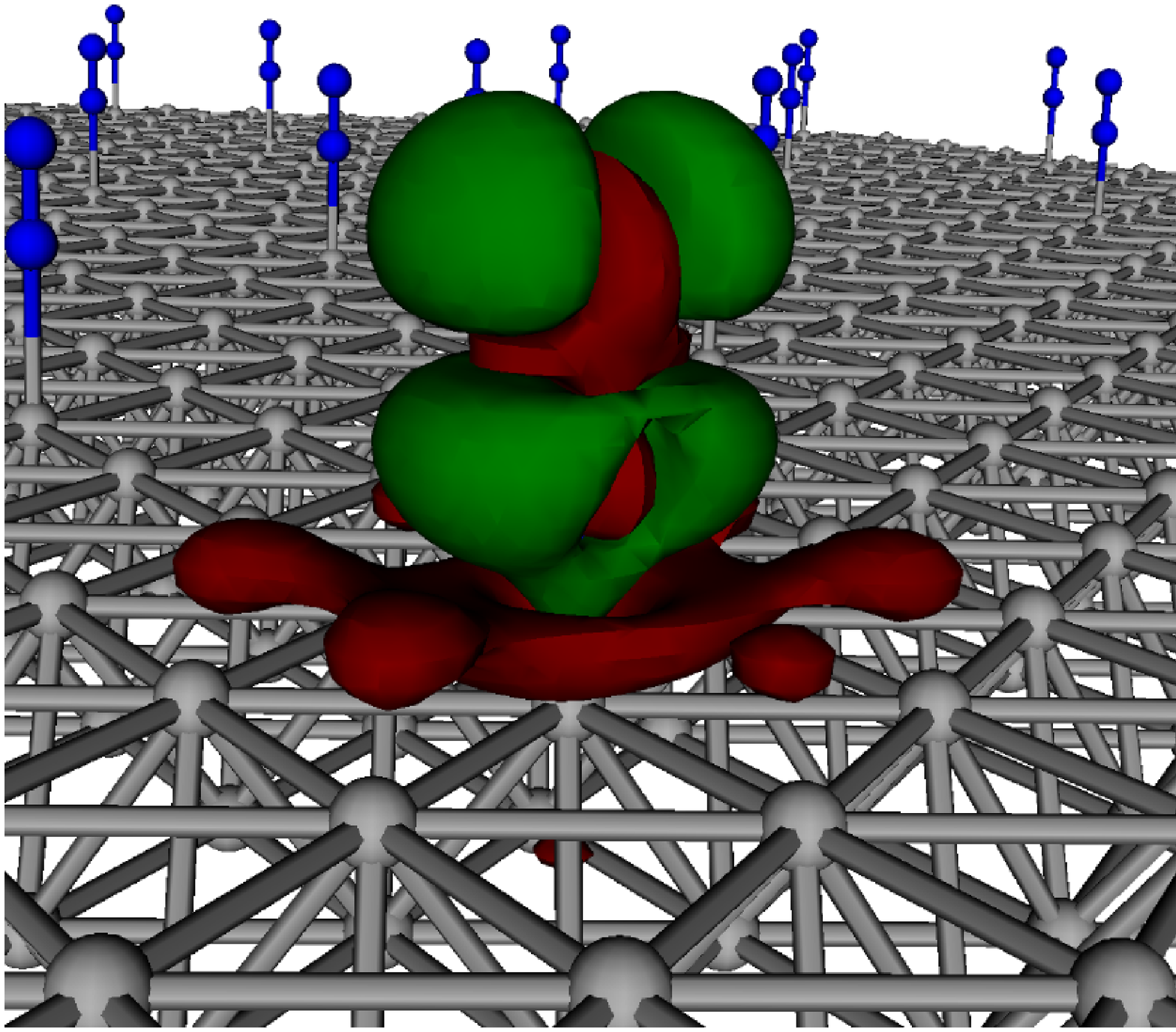}
\caption{(color) The change in charge distribution due to the excitation. Green: more charge (0.01 a.u.\ contour), red: less charge (-0.01 a.u.\ contour). The 4 figures are for 4 different surface unit cells: (1,1), (2,1), (2,2) and (4,2). Grey atoms are ruthenium and blue atoms are nitrogen. The periodic images of the atoms are also shown, whereas the density changes are only shown in one unit cell.}
\label{fig:vtk_RuN2_area}
\end{center}
\end{figure}

The resonance energy is converged to within 0.1 eV at a surface unit cell of (2,2). The rather large variation in energy for smaller unit cells is probably due to dipole interactions between periodic images. This is confirmed by a simple estimation of the dipol-dipol interaction energies. The resonance energy is not influenced significantly by the number of layers in the ruthenium, indicating that the charge redistribution only occurs very near to the surface. That the charge redistribution is local is confirmed by Fig.\ \ref{fig:vtk_RuN2_area}, which shows the change in charge between the resonance calculation and the ground state calculation for 4 different surface unit cells.
For the larger unit cells, where the resonance energy has converged, a clearly localized image charge is seen below the nitrogen molecule and above the first layer of ruthenium atoms. The area with extra charge clearly resembles the $2\pi$ orbital of nitrogen, indicating that the $2\pi$ orbital is well represented by the linear expansion of KS orbitals. Fig.\ \ref{fig:vtk_RuN2_area} also reveals that some charge is redistributed within the molecule.

In order to get an estimate of the size of the charge redistribution we also performed Bader decomposition\cite{bader_book,bader} on the density found in the ground state calculation and the resonance calculation. The two bottom panels on Fig.\ \ref{fig:bader_convergence} show the extra charge assigned to the nitrogen molecule in the resonance calculation compared to the ground state calculation as a function of system size. The converged value is close to 0.5 electron charge, ie.\ only half of the electron is placed on the nitrogen molecule according to the Bader decomposition. This discrepancy could either be due to the ambiguity in the way one chooses to assign charge to the atoms or a more physical effect of charge going back into the surface when extra charge is placed on the molecule. The former reason is very likely, since the image charge is located very close to the molecule.

\begin{figure}[t]
\begin{center}
\includegraphics[width=0.4\textwidth]{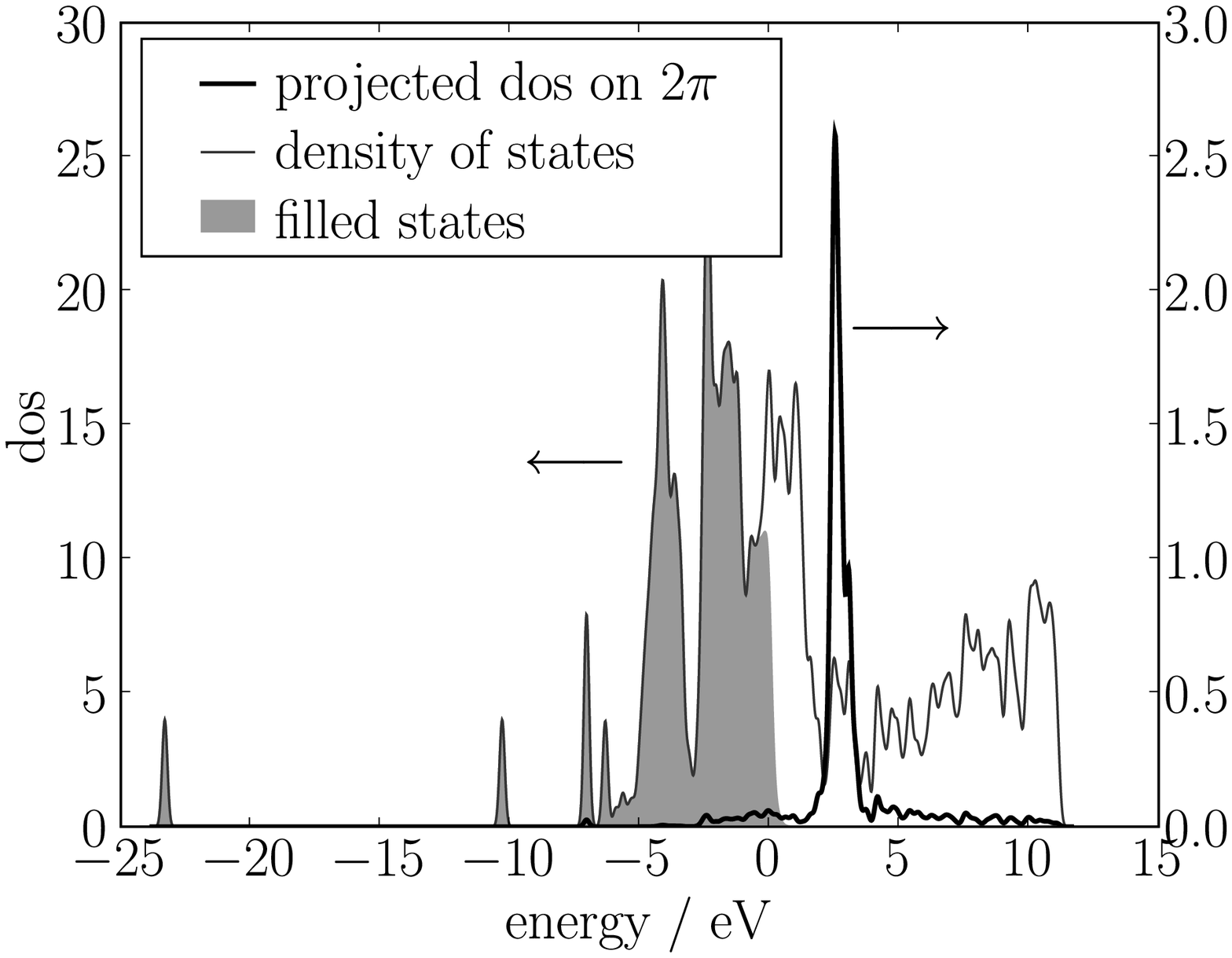}\\
\includegraphics[width=0.4\textwidth]{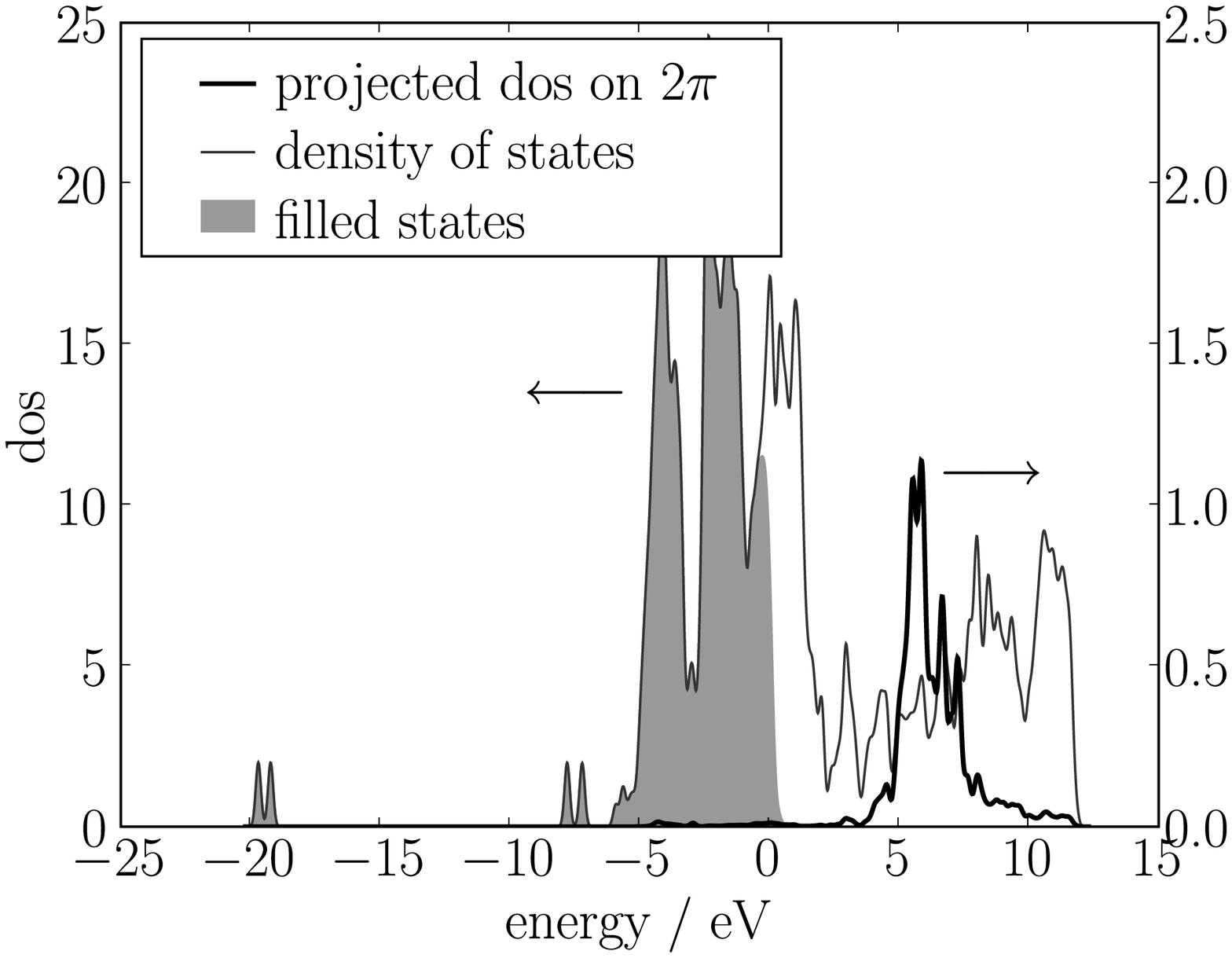}
\caption{The density of states for a N$_2$ molecule on a ruthenium slab and the projected density of states on the $2\pi$ orbital of the N$_2$ molecule. Top: Ground state calculation. Bottom: Resonance calculation.}\label{fig:dos}
\end{center}
\end{figure}

\begin{figure}[t]
\begin{center}
\includegraphics[width=0.4\textwidth]{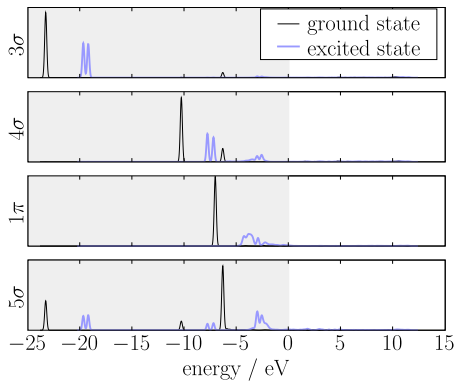}
\caption{(color online) Projected density of states (PDOS) on the $3\sigma$, $4\sigma$, $1\pi$ and $5\sigma$ orbitals of a N$_2$ molecule sitting on a ruthenium slab. The PDOSs are plotted for both the ground state calculation and the resonance calculation. The grey area indicates energies below the Fermi level.}\label{fig:pdos}
\end{center}
\end{figure}

In order to investigate the effect of charge going back into the surface we start by considering the $2\pi$ orbital itself. Fig.\ \ref{fig:dos} shows the density of KS states and the projected density of states (PDOS) for the $2\pi$ orbital for the ground state calculation and the resonance calculation.
In the ground state calculation a part of the long tail of the PDOS goes below the Fermi energy, ie.\ a small part of the $2\pi$ orbital is occupied here. In the resonance calculation the PDOS has moved upward in energy such that the tail no longer goes below the Fermi energy, ie.\ some charge goes back into the surface as charge is placed on the molecule. Similar effects are seen for the other molecular orbitals as visualized on Fig.\ \ref{fig:pdos}, which shows the PDOS for the $3\sigma$, $4\sigma$, $1\pi$ and  $5\sigma$ orbitals. 
Again it is seen that all the PDOSs are shifted up in energy as more charge is placed on the molecule. Almost the entire PDOSs are still under the Fermi level, but small ripples can be seen above the Fermi level, also contributing to the amount of charge going back into the surface.

This backtransfer of charge is not an unwanted effect, since we try to model the long-lived resonance state, ie.\ the resonably localized peak in the PDOS on Fig.\ \ref{fig:dos}. The backtransfer of charge is due to some on the energy scale very delocalized bands, indicating a much shorter lifetime, ie.\ the backtransfer is expected to happen on a much shorter time scale than the decay of the resonance. It is however clear from Figs.\ \ref{fig:dos} and \ref{fig:pdos} that the charge backtransfer in this case is far from the 0.5 electron indicated by the Bader decomposition. We then conclude that the main part of the discrepancy in this situation can be assigned to the ambiguity in the way charge is assigned to the different atoms. We also find that one gets significant different results by assigning charge in a different manner, for example by dividing the charge by a flat plane midway between the surface and the molecule.

\subsection{Comparison with inverse-photoemission spectroscopy experiments}

In Table \ref{table:surf_test} we have tested the linear expansion $\Delta$SCF method against inverse-photoemission spectroscopy (IPES) measurements and compared the results to spatially constrained DFT and ordinary $\Delta$SCF calculations.
\begin{table}[t]
\begin{center}
\caption{Comparison of the 2$\pi$ resonance energies for different diatomic molecules on different surfaces found by spatially constrained DFT, ordinary $\Delta$SCF, our modified $\Delta$SCF and experiments. The experimental results have been obtained from inverse-photoemission spectroscopy (IPES) measurements: $^a$Johnson and Hulbert,\cite{ipes} $^b$Reimer et al,\cite{reimer_fink_kuppers_1987} $^c$Reimer et al,\cite{reimer_fink_kuppers_1988} $^d$Rogozik and Dose\cite{rogozik_dose_1986}.  All energies are in eV. We have not included lrTDDFT calculations, since it is not applicable to periodic systems.}\label{table:surf_test}
\begin{tabular}{lcccr}
\hline\hline
System & Constrained & $\Delta$SCF & $\Delta$SCF & Experiment \\
       & DFT         & (orig.)     & (this work) &     \\
\hline
N$_2$ on Ni(001) &  2.2 &  3.5 &  4.0  & 4.4$^a$ \\
CO on Ni(001)    &  2.2 &  3.2 &  4.2  & 4.0$^a$/4.5$^b$ \\
NO on Ni(001)    &  2.2 &  0.6 &  1.4  & 1.6$^a$/1.5$^c$ \\
CO on Ni(111)    &  2.8 &  4.3 &  4.4  & 4.4$^c$ \\
NO on Ni(111)    &  2.7 &  0.5 &  1.4  & 1.5$^b$ \\
CO on Pd(111)    &  4.6 &  4.1 &  4.9  & 4.7$^d$ \\
CO on Pd step    &  2.8 &  3.2 &  4.5  & 4.0$^d$ \\
\hline\hline
\end{tabular}
\end{center}
\end{table}
The modified $\Delta$SCF values are all calculated in exactly the same manner as for N$_2$ on ruthenium in the previous section. In all cases the molecules sit on-top, and all surface atoms were kept fixed during the minimization of the molecular degrees of freedom. For the Ni (001) surface we used 3 atomic layers, for the Ni (111) and Pd surfaces we used 2 atomic layers. The positions of the molecules in their minimized position is given in table \ref{table:atomic_pos}.
\begin{table}[t]
\begin{center}
\caption{The positions of the molecules in the systems from table \ref{table:surf_test}. All positions are with relative to the closest surface atom. The z-direction is normal to the surface. At the Pd step the CO molecule is tilted over the step, which is the reason for the composant in the y-direction. All number are i Angstrom}\label{table:atomic_pos}
\begin{tabular}{llll}
\hline\hline
Surface & Molecule & Pos. of 1. atoms & Pos. of 2. atom \\
\hline
Ni(001) & N$_2$& N: (0,0,1.638) & N: (0,0,2.798) \\
        & CO   & C: (0,0,1.456) & O: (0,0,2.621) \\
        & NO   & N: (0,0,1.404) & O: (0,0,2.580) \\
\hline
Ni(111) & CO   & C: (0,0,1.774) & O: (0,0,2.941) \\
        & NO   & N: (0,0,1.758) & O: (0,0,2.935) \\
\hline
Pd(111) & CO   & C: (0,0,1.904) & O: (0,0,3.064) \\
Pd step & CO   & C: (0,0.586,1.801) & O: (0,0.844,2.934) \\
\hline\hline
\end{tabular}
\end{center}
\end{table}
All resonance energies are vertical from the minimum of the ground state PES. The relevant resonance for all the considered systems is the $2\pi$ resonance. 

The spatially constrained DFT method was suggested by Wu and van Voorhis.\cite{wu_voorhis_2005,wu_voorhis_2006} In the calculations we perform here we divide the space into two areas divided by the flat plane mid between the surface and the lowest atom in the molecule. We the apply a potential, $V=V_0\cdot(1+\exp(\tfrac{z_0-z}{\Delta z}))^{-1}$, with $\Delta z=0.2$ {\AA} and $z_0$ being the z-value of the dividing plane. $V_0$ is varied until an extra electron is placed on the molecules side of the dividing plane compared to the unconstrained calculation. The energy is then calculated as described by Wu and van Voorhis.\cite{wu_voorhis_2005,wu_voorhis_2006} The results using the original $\Delta$SCF method have all been obtained by forcing an electron in the KS orbital with the largest overlap with the 2$\pi$ orbital.

The results obtained with our proposed modification of the $\Delta$SCF method is seen to agree quite well with the experimental results, better than the spatially constrained DFT and the original $\Delta$SCF method. All the results obtained by the original $\Delta$SCF approach lie too low, which is due to the fact that the large hybridization of the molecular orbitals makes it impossible to place sufficient charge on the molecule. However, a significant problem with this method is that PESs often become discontinuous if one chooses to occupy the KS orbital with the largest overlap with the molecular orbital, since this can be different orbitals at different configurations. 

The major problem with the spatially constrained DFT method seems to be that it in some cases is a too strict criteria to force an extra electron on the molecule, which reflects itself in similar resonance energies for CO and NO. We find that the backtransfer of charge discussed in the last section is significant for adsorbed NO and essential to obtain the resonance energies we find with the modified $\Delta$SCF method. This indicates that the spatially constrained DFT approach is more suited for systems with a smaller coupling than one has on the metallic surfaces considered here. The good agreement between our modified $\Delta$SCF method and experiments indicate that this method is preferable for these kinds of systems and that the backtransfer effect is indeed physically reasonable.

\subsection{Potential energy surfaces for N$_2$ on ruthenium}\label{sec:pesRuN2}

\begin{figure}[t]
\begin{center}
\includegraphics[width=0.4\textwidth]{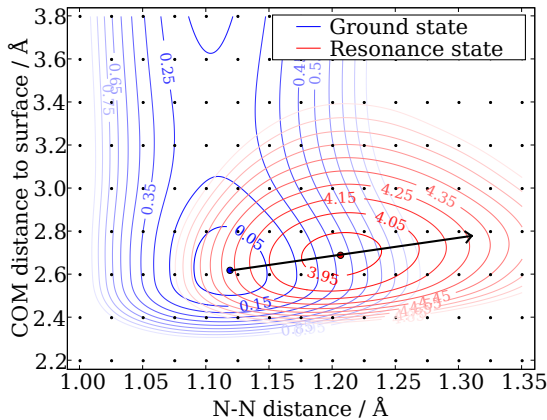}
\caption{(color) Potential energy surfaces (PES) for a nitrogen molecule on a close-packed ruthenium surface in the ground state and the $2\pi_y$ resonance as a function of the distance between the two nitrogen atoms and the distance from the surface to the center of mass of the nitrogen molecule. The energies are in eV. The small dots represents the points where the energy has been calculated in order to generate the surfaces. The black arrow represent a possible trajectory of the system in the resonance state (see text).}\label{fig:pes}
\end{center}
\end{figure}

On Fig.\ \ref{fig:pes} we have mapped out a part of the potential energy surfaces (PES) for a nitrogen molecule on a ruthenium (0001) surface in the ground state and the $2\pi_y$ resonance. We limit ourselves to two dimensions, which at least is reasonable in the ground state, since here it is well known that the molecule sits vertically on an on-top site. In the resonance state we have tried to rotate the molecule a small angle around the surface atom in the x- and y-direction at several points on the PES. In all cases this leads to an increase in energy, ie. it also seems reasonable to stay within the two dimensions in the resonance state. Here we will only apply the PES to a simple estimate of the possible energy transfer into molecular motion from an electron hitting the resonance. For a more detailed analysis it is necessary to include other dimensions.

The ground state PES looks as expected, with a small barrier for desorption and a local minimum corresponding to the adsorption configuration. The resonance PES has a shifted minimum, which indicates that an electron hitting this resonance could induce molecular motion, since a sudden shift between the PESs would leave the system far away from the minimum, such that the atoms would start to move. The maximum possible energy gain assuming classical ion dynamics from a single electron hitting the resonance can be roughly estimated by following the black arrow on Fig.\ \ref{fig:pes}. The system is most likely situated at the local minimum of the ground state PES when the electron hits the resonance. The black arrow shows a possible trajectory of the system in the resonance state until the resonance decays and the system returns to the ground state PES. The potential energy after the electron event in this optimal situation is approximately 1.5 eV higher than before the event. This is seen to be more than enough to desorb the molecule. A more detailed analysis involving calculations of the possible vibrational excitations and the probabilities of exciting them will be the topic of a future publication. Such an analysis will have to take all six degrees of freedom of the molecule into account.

The PESs show that the center of mass is shifted away from the surface when the resonance is occupied. This may seem counter intuitive since the charged molecule is attracted to the generated image charge in the surface. However, the resonance weakens the bond between the nitrogen atoms, such that the distance between them increases, which shifts the center of mass outwards as the lower atom is not free to move closer to the surface. This effect is more significant than the decrease in the ruthenium-nitrogen distance due to the mentioned image charge effect.

\section{Small molecules}

In the following we present some small tests performed on N$_2$ and CO. These small systems have the advantage that they make it possible to compare to more accurate linear response time-dependent density functional theory (lrTDDFT) calculations. When possible we also compare to experiments. The only advantage of our modified $\Delta$SCF compared to ordinary $\Delta$SCF for these molecules is the possibility of handling degenerate states without getting convergence problems, ie.\ the following should be viewed as a test of the $\Delta$SCF approach rather than a test of our modification. We are especially interested in confirming the ability to predict the shift of the minimum when going from the ground state PES to the excited state PES, which we in section \ref{sec:pesRuN2} argued is very important when considering molecular motion induced by an electron hitting a molecular resonance.

\subsection{Excitation energies}

We have used the linear expansion $\Delta$SCF in combination with the multiplet sum method\cite{multiplet} to calculate excitation energies for different excitations in the N$_2$ and CO molecule. The results are presented in table \ref{table:N2_e} and \ref{table:CO_e} respectively.
\begin{table}[t]
\begin{center}
\caption{Vertical excitation energies for the N$_2$ molecule taken from the minimum energy configuration of the ground state. $^a$KS eigenvalue differences, $^b$linear response calculations taken from Grabo et al.\cite{lrtddftN2}, $^c$Computed by Oddershede et al\cite{expN2_a} using the spectroscopic constants of Huber and Herzberg.\cite{expN2_b} All theoretical results are obtained using LDA as the xc-potential (and ALDA for the xc-kernel in the lrTDDFT calculations).}\label{table:N2_e}
\begin{tabular}{lcrrrrr}
\hline\hline
\multirow{2}{*}{State} & \multirow{2}{*}{Transition} & \multirow{2}{*}{$\Delta\epsilon_{KS}\;^a$} & TDDFT$^b$ & $\Delta$SCF & $\Delta$SCF & \multirow{2}{*}{Exp.$^c$} \\
      &            &                          &(ALDA)       & (LDA)       & (RPBE)      &          \\
\hline
$a^1\Pi$ & \multirow{2}{*}{$5\sigma \rightarrow 2\pi$} & \multirow{2}{*}{8.16} &  9.23 &  8.75 &  8.58 & 9.31 \\
$B^3\Pi$ &                                             &                       &  7.62 &  7.55 &  7.52 & 8.04 \\
\multicolumn{3}{c}{singlet-triplet splitting:}                                 &  1.61 &  1.20 &  1.06 & 1.27 \\
\hline
$w^1\Delta$ & \multirow{2}{*}{$1\pi \rightarrow 2\pi$} & \multirow{2}{*}{9.63} & 10.27 & 10.50 & 10.52 &10.27 \\
$W^3\Delta$ &                                          &                       &  8.91 &  8.94 &  8.79 & 8.88 \\
\multicolumn{3}{c}{singlet-triplet splitting:}                                 &  1.36 &  1.56 &  1.73 & 1.39 \\
\hline
$o^1\Pi$ & \multirow{2}{*}{$4\sigma \rightarrow 2\pi$} &\multirow{2}{*}{11.21} & 13.87 & 11.97 & 12.40 &13.63 \\
$C^3\Pi$ &                                             &                       & 10.44 & 10.37 & 10.61 &11.19 \\
\multicolumn{3}{c}{singlet-triplet splitting:}                                 &  3.43 &  1.60 &  1.79 & 2.44 \\
\hline\hline
\end{tabular}
\end{center}
\end{table}
\begin{table}[t]
\begin{center}
\caption{Vertical excitation energies for the CO molecule taken from the minimum energy configuration of the ground state. $^a$KS eigenvalue differences, $^b$linear response calculations taken from Gross et al.\cite{lrtddftCO}, $^c$Computed by Nielsen et al.\cite{expCO} All theoretical results are obtained using LDA as the xc-potential (and ALDA for the xc-kernel in the lrTDDFT calculations).}\label{table:CO_e}
\begin{tabular}{lcrrrrr}
\hline\hline
\multirow{2}{*}{State} & \multirow{2}{*}{Transition} & \multirow{2}{*}{$\Delta\epsilon_{KS}\;^a$} & TDDFT$^b$ & $\Delta$SCF & $\Delta$SCF & \multirow{2}{*}{Exp.$^c$} \\
      &            &                          &(ALDA)       & (LDA)       & (RPBE)      &          \\
\hline
$A^1\Pi$ & \multirow{2}{*}{$5\sigma \rightarrow 2\pi$}  & \multirow{2}{*}{6.87} &  8.44 &  7.84 &  7.81 & 8.51 \\
$a^3\Pi$ &                                              &                       &  6.02 &  6.09 &  6.02 & 6.32 \\
\multicolumn{3}{c}{singlet-triplet splitting:}                                  &  2.42 &  1.75 &  1.79 & 2.19 \\
\hline
$D^1\Delta$ & \multirow{2}{*}{$1\pi \rightarrow 2\pi$}  & \multirow{2}{*}{9.87} & 10.36 & 10.82 & 10.73 &10.23 \\
$d^3\Delta$ &                                           &                       &  9.24 &  9.72 &  9.55 & 9.36 \\
\multicolumn{3}{c}{singlet-triplet splitting:}                                  &  1.12 &  1.10 &  1.18 & 0.87 \\
\hline
$C^1\Pi$ & \multirow{2}{*}{$4\sigma \rightarrow 2\pi$}  & \multirow{2}{*}{11.94}&   -   & 13.15 & 13.09 &  -   \\
$c^3\Pi$ &                                              &                       & 11.43 & 12.26 & 12.09 &11.55 \\
\multicolumn{3}{c}{singlet-triplet splitting:}                                  &   -   &  0.89 &  1.00 &  -   \\
\hline\hline
\end{tabular}
\end{center}
\end{table}
The $4\sigma$ and $5\sigma$ states are both represented by a single KS orbital. The $1\pi$ and $2\pi$ states are both double degenerate, so they are both represented as a linear combination of two KS orbitals: $\vert\pi\rangle=\tfrac{1}{\sqrt{2}}\vert\pi_{KS,a}\rangle+i\tfrac{1}{\sqrt{2}}\vert\pi_{KS,b}\rangle$, where $\vert\pi_{KS,a}\rangle$ and $\vert\pi_{KS,b}\rangle$ are the two degenerate KS orbitals. The imaginary unit, $i$, has been included in order to get the correct angular momentum of the excited states ($\Pi$ and $\Delta$). This would not be possible using traditional $\Delta$SCF, where one only has the freedom to change occupation numbers of the KS states. Due to the rotational symmetry of the density found from these states the calculations do not suffer from any convergence difficulties. That is not the case if one just occupies one of the degenerate KS orbitals. Only the $\Delta$ states are included in the $1\pi \rightarrow 2\pi$ transitions in Tables \ref{table:N2_e} and \ref{table:CO_e}, since the $\Sigma$ states cannot be estimated by the multiplet sum method.\cite{multiplet} This is, however, not a problem for the kinds of systems for which this method is intended, such as molecules on surfaces where high-accuracy alternatives are still lacking.

In general the excitation energies found by the linear expansion $\Delta$SCF method look quite good for the low-lying excitations. The accuracy is only slightly worse than that of lrTDDFT and significantly better than just taking KS eigenvalue differences. The singlet triplet splittings are also rather close to the experimental values. The method however seems to struggle a bit more in the higher lying $4\sigma\rightarrow2\pi$ transitions. This could indicate that the method should only be applied to low lying excitations. Changing the exchange-correlation functional from LDA to RPBE does not affect the accuracy significantly although a small tendency towards better performance is seen for the higher lying excitations. We note, however, that the intended application of $\Delta$SCF do not include simple diatomic molecules, where more accurate quantum chemical methods are available.

\subsection{Excited potential energy surfaces}

\begin{figure}[t]
\begin{center}
\includegraphics[width=0.4\textwidth]{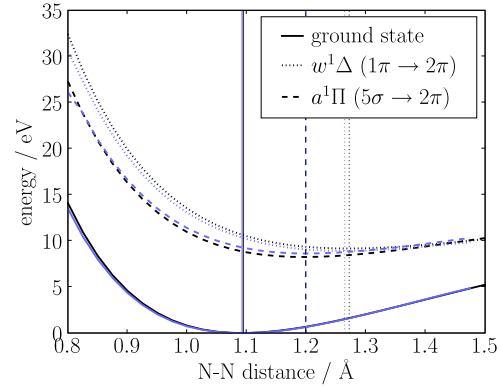}
\caption{(color online) The energy as a function of bond length for the N$_2$ molecule in the ground state and two excited states. The black lines correspond to $\Delta$SCF calculations, the grey (online: light blue) lines correspond to linear response calculations. The linear response calculations have been made using \texttt{Octopus}.\cite{octopus,octopus_paper} The vertical lines indicate the positions of the minima.}\label{fig:e_vs_bl_N2}
\end{center}
\end{figure}

The shapes of the potential energy surfaces can in some cases be more important than the exact height of them, ie.\ a constant error is not so critical. This is for example the case when considering chemistry induced by hot electrons.\cite{gadzuk_1996,Saalfrank_DIET_DIMET} In order to get an idea of the accuracy with which the linear expansion $\Delta$SCF method reproduces correct shapes of potential energy surfaces we have calculated the potential energy surfaces for the ground state and two excited states in the N$_2$ molecule. These are plotted in Fig.\ \ref{fig:e_vs_bl_N2} together with results from lrTDDFT calculations.

The small difference between the two ground state curves are due to the fact that they have been calculated with two different codes. Both codes are realspace codes, but \texttt{gpaw} uses the PAW formalism to represent the core electrons whereas \texttt{Octopus} uses norm-conserving pseudopotentials. The calculations have been made with the same grid spacing and unit cell size, and with the same exchange-correlation potential (LDA/ALDA).

The shapes of the potential energy surfaces found from the two different methods are seen to be very similar. Especially the predicted positions of the minima are seen to agree very well. The shifting of the minima towards larger bond lengths is also the expected behavior, since an electron is moved from an bonding orbital to an anti-bonding orbital. When going to bond lengths beyond 2 {\AA} we start having problems with convergence problems in the $\Delta$SCF calculations, since the $2\pi$ orbital ceases to exist. This is not a problem we have encountered in the systems with a molecule on a surface.

The good agreement between $\Delta$SCF and lrTDDFT probably reflects that $\Delta$SCF and ignoring the history dependence of the exchange-correlation potential in TDDFT are related approximations. For example, the density obtained in $\Delta$SCF would be stationary if evolved in time with TDDFT.

\subsection{Gradients}\label{sec:gradients_calc}

\begin{figure}[t]
\begin{center}
\includegraphics[width=0.4\textwidth]{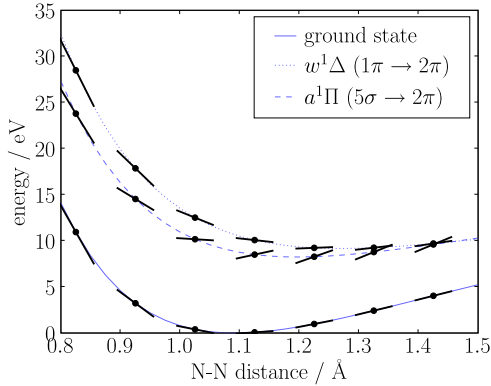}
\caption{(color online) The energy as a function of bond length for the N$_2$ molecule in the ground state and two excited states. The short thick lines indicate the size of the gradients.}\label{fig:f_vs_bl_N2}
\end{center}
\end{figure}

As mentioned in section \ref{sec:method_forces} the Hellman-Feynman theorem does not apply in the linear expansion $\Delta$SCF method. This is verified by the calculations shown on Fig.\ \ref{fig:f_vs_bl_N2}. Here the energy of the ground state and two excited states in the N$_2$ molecule is plotted as a function of the bond length. The short thick lines indicate the gradient given by calculated Hellman-Feynman forces.
For the ground state the agreement is as expected perfect, but for the excited states there is a clear mismatch. Unfortunately this implies that it is computationally heavy to do dynamics or minimizations in the excited states.

\section{Summary}

We have extended the $\Delta$SCF method of calculating excitation energies by allowing excited electrons to occupy linear combinations of KS states instead of just single KS states. This solves the problems encountered for molecules near surfaces, where the molecular orbitals hybridize, such that none of the KS orbitals can be used to represent an extra electron placed on the molecule. The method has been implemented in \texttt{gpaw}\cite{gpaw_article,gpaw} and applied to several systems.

From calculated potential energy surfaces of N$_2$ on a ruthenium surface we concluded that an electron hitting the 2$\pi$ resonance in this system can induce molecular dynamics due to the different positions of the minima of the ground state PES and the resonance PES. Through a simple analysis we found that one electron can optimally place 1.5 eV in the atomic motion, more than enough to desorb the molecule.

We find good agreement between the model and inverse photo-emission experiments for several diatomic molecules on different metallic surfaces. For the considered systems we find significantly better agreement with experiments using the modified $\Delta$SCF method compared to spatially constrained DFT and traditional $\Delta$SCF.

Finally we applied the method to N$_2$ and CO in their gas phases we found that excitation energies are estimated with quite good accuracy for the lower lying excitations, comparable to that of TDDFT. Especially the shape of the potential energy surfaces and the positions of the minima agree well with TDDFT results.

\begin{acknowledgments}
The authors wish to thank Anders Hellman for fruitful discussions.  The
Center for Individual Nanoparticle Functionality (CINF) is sponsored
by the Danish National Research Foundation.  This work was supported
by the Danish Center for Scientific Computing.
\end{acknowledgments}

\appendix*

\section{Overlaps using PAW pseudo wavefunction projections}\label{app:paw}

The Projector Augmented Wave (PAW)\cite{paw1} method utilizes that one can transform single-particle wavefunctions $|\psi_n\rangle$ oscillating wildly near the atom core (all-electron wavefunctions), into smooth well-behaved wavefunctions $|\tilde\psi_n\rangle$ (pseudo wavefunctions) which are identical to the all-electron wavefunctions outside some augmentation sphere. The idea is to expand the pseudo wavefunction inside the augmentation sphere on a basis of smooth continuations $|\tilde\phi_i^a\rangle$ of partial waves $|\phi_i^a\rangle$ centered on atom $a$. The transformation is
\begin{align}\label{PAW}
|\psi_n\rangle=|\tilde\psi_n\rangle+\sum_{i,a}\Big(|\phi_i^a\rangle-|\tilde\phi_i^a\rangle\Big)\langle\tilde p_i^a|\tilde\psi_n\rangle,
\end{align}
where the projector functions $|\tilde p_i^a\rangle$ inside the augmentation sphere $a$ fulfills
\begin{align}\label{dual}
\sum_i|\tilde p_i^a\rangle\langle\tilde\phi_i^a|=1,\qquad\langle\tilde p_i^a|\tilde\phi_j^a\rangle=\delta_{ij},\qquad|\mathbf{r}-\mathbf{R}^a|<r^a_c.
\end{align}

Suppose we have an atom adsorbed on a metal surface and we wish to perform a $\Delta$SCF calculation where a certain atomic orbital $|a\rangle$ is kept occupied during the calculation. If the orbital is hybridized with the metal states we need to find the linear combination which constitutes the orbital. This can always be done if a sufficient number of unoccupied KS orbitals is included in the calculation
\begin{align}\label{linear}
|i\rangle=\sum_nc_{ni}|\psi_n\rangle, \qquad c_{ni}=\langle\psi_n|i\rangle
\end{align}
Since the partial waves are typically chosen as atomic orbitals we just need to consider the quantity
\begin{eqnarray}
\langle\psi_n|\phi_i^a\rangle&=&\langle\tilde\psi_n|\phi_i^a\rangle+\sum_{j,a'}\langle\tilde\psi_n|\tilde p_j^{a'}\rangle\Big(\langle\phi^{a'}_j|\phi_i^a\rangle-\langle\tilde\phi^{a'}_j|\phi_i^a\rangle\Big)\nonumber\\
&\approx&\langle\tilde\psi_n|\tilde p_i^a\rangle.
\end{eqnarray}
If we were just considering a single atom, the last equality would be exact inside the augmentation sphere since the partial waves would then be orthogonal and the pseudo partial waves are dual to the projectors in Eq.\ \eqref{dual}. When more than one atom is present there is corrections due to overlap of partial waves from neighboring atoms and non-completeness of projectors/pseudo partial waves between the augmentation spheres. However using $\langle\tilde\psi_n|\tilde p_i^a\rangle$ is a quick and efficient way of obtaining the linear combination, since these quantities are calculated in each step of the self-consistence cycle anyway. The method can then be extended to molecular orbitals by taking the relevant linear combinations of $\langle\tilde\psi_n|\tilde p_i^a\rangle$.


\begin{thebibliography}{49}
\bibitem[{e_t()}]{e_temp_footnote}
\bibinfo{note}{Often an electronic temperature is introduced to improve
  convergence, such that the electrons are distributed by a Fermi-Dirac
  distribution. This can also be done in the proposed method.}
\end{thebibliography}

\end{document}